\documentclass{pasa}

\usepackage{graphicx}

\title{Interpretable Faraday Complexity Classification}

\author[M. J. Alger et al.]
{
  M.~J.~Alger$^{1,2}$\thanks{matthew.alger@gmail.com},
  J.~D.~Livingston$^1$,
  N.~M.~McClure-Griffiths$^1$,
  J.~L.~Nabaglo,
  O.~I.~Wong$^{3,4,5}$,
  C.~S.~Ong$^{2,6}$
\affil{$^1$Research School of Astronomy and Astrophysics, The Australian National University, Canberra, ACT 2611, Australia}%
\affil{$^2$Data61, CSIRO, Canberra, ACT 2601, Australia}%
\affil{$^3$CSIRO Astronomy \& Space Science, PO Box 1130, Bentley, WA 6102, Australia}%
\affil{$^4$ICRAR-M468, University of Western Australia, Crawley, WA 6009, Australia}%
\affil{$^5$ARC Centre of Excellence for All Sky Astrophysics in 3 Dimensions (ASTRO 3D), Australia}%
\affil{$^6$Research School of Computer Science, The Australian National University, Canberra, ACT 2601, Australia}%
}%

\jid{PASA}
\doi{10.1017/pas.\the\year.xxx}
\jyear{\the\year}

\usepackage{aas_macros}
\usepackage{hyperref}
\hypersetup{colorlinks,citecolor=blue,linkcolor=blue,urlcolor=blue}

\usepackage{cleveref}
\usepackage{upgreek}
\usepackage{bm}
\renewcommand{\vec}{\bm}
\newcommand{\citeneeded}[1]{{\color{cyan} (citation needed)}}
\usepackage{mathtools}
\DeclarePairedDelimiterX{\infdivx}[2]{(}{)}{%
  #1\;\delimsize\|\;#2%
}

\usepackage{subcaption}

\usepackage{multirow}

\begin{document}

\begin{frontmatter}
\maketitle

\begin{abstract}
  Faraday complexity describes whether a spectropolarimetric observation has simple or complex magnetic structure. Quickly determining the Faraday complexity of a spectropolarimetric observation is important for processing large, polarised radio surveys. Finding simple sources lets us build rotation measure grids, and finding complex sources lets us follow these sources up with slower analysis techniques or further observations. We introduce five features that can be used to train simple, interpretable machine learning classifiers for estimating Faraday complexity. We train logistic regression and extreme gradient boosted tree classifiers on simulated polarised spectra using our features, analyse their behaviour, and demonstrate our features are effective for both simulated and real data. This is the first application of machine learning methods to real spectropolarimetry data. With 95 per cent accuracy on simulated ASKAP data and 90 per cent accuracy on simulated ATCA data, our method performs comparably to state-of-the-art convolutional neural networks while being simpler and easier to interpret. Logistic regression trained with our features behaves sensibly on real data and its outputs are useful for sorting polarised sources by apparent Faraday complexity.
\end{abstract}

\end{frontmatter}

\renewcommand{\sectionautorefname}{Section}%
\renewcommand{\subsectionautorefname}{Section}%
\renewcommand{\subsubsectionautorefname}{Section}%

\section{Introduction}
\label{sec:intro}

  As polarised radiation from distant galaxies makes its way to us, magnetised plasma along the way can cause the polarisation angle to change due to the Faraday effect. The amount of rotation depends on the squared wavelength of the radiation, and the rotation per squared wavelength is called the Faraday depth. Multiple Faraday depths may exist along one line-of-sight, and if a polarised source is observed at multiple wavelengths then these multiple depths can be disentangled. This can provide insight into the polarised structure of the source or the intervening medium.

  Faraday rotation measure synthesis (RM synthesis) is a technique for decomposing a spectropolarimetric observation into flux at its Faraday depths $\phi$, the resulting distribution of depths being called a `Faraday dispersion function' (FDF) or a `Faraday spectrum'. It was introduced by \citet{brentjens_faraday_2005} as a way to rapidly and reliably analyse the polarisation structure of complex and high-Faraday depth polarised observations.

  A `Faraday simple' observation is one for which there is only one Faraday depth, and in this simple case the Faraday depth is also known as a `rotation measure' (RM). All Faraday simple observations can be modelled as a polarised source with a thermal plasma of constant electron density and magnetic field \citep[a `Faraday screen';][]{brentjens_faraday_2005,anderson_broadband_2015} between the observer and the source. A `Faraday complex' observation is one which is not Faraday simple, and may differ from a Faraday simple source due to plasma emission or composition of multiple screens \citep{brentjens_faraday_2005}. The complexity of a source tells us important details about the polarised structure of the source and along the line-of-sight, such as whether the intervening medium emits polarised radiation, or whether there are turbulent magnetic fields or different electron densities in the neighbourhood. The complexity of nearby sources taken together can tell us about the magneto-ionic structure of the galactic and intergalactic medium between the sources and us as observers. \citet{osullivan_broad-band_2017} show examples of simple and complex sources, and \autoref{fig:simple-fdf} and \autoref{fig:complex-fdf} show an example of a simulated simple and complex FDF respectively.

  Identifying when an observation is Faraday complex is an important problem in polarised surveys \citep{sun15comparison}, and with current surveys such as the Polarised Sky Survey of the Universe's Magnetism (POSSUM) larger than ever before, methods that can quickly characterise Faraday complexity en masse are increasingly useful. Being able to identify which sources are simple lets us produce a reliable rotation measure grid from background sources, and being able to identify which sources might be complex allows us to find sources to follow-up with slower polarisation analysis methods that may require manual oversight, such as QU fitting \citep[as seen in e.g.][]{miyashita19qu,osullivan_broad-band_2017}. In this paper, we introduce five simple, interpretable features representing polarised spectra, use these features to train machine learning classifiers to identify Faraday complexity, and demonstrate their effectiveness on real and simulated data. We construct our features by comparing observed polarised sources to idealised polarised sources. The features are intuitive and can be estimated from real FDFs.

  \autoref{sec:background} provides a background to our work, including a summary of prior work and our assumptions on FDFs. \autoref{sec:approach} describes our approach to the Faraday complexity problem. \autoref{sec:experiment-classification} explains how we trained and evaluated our method. Finally, \autoref{sec:discussion} discusses these results.

\section{Faraday Complexity}
\label{sec:background}

    Faraday complexity is an observational property of a source: if multiple Faraday depths are observed within the same apparent source (e.g. due to multiple lines-of-sight being combined within a beam), then the source is complex. A source composed of multiple Faraday screens may produce observations consistent with many models \citep{sun15comparison}, including simple sources, so there is some overlap between simple and complex sources. Faraday thickness is also a source of Faraday complexity: when the intervening medium between a polarised source and the observer also emits polarised light, the FDF cannot be characterised by a simple Faraday screen. As discussed in \autoref{sec:fdfs} we defer Faraday thick sources to future work. In this section we summarise existing methods of Faraday complexity estimation and explain our assumptions and model of simple and complex polarised FDFs.

  \subsection{Prior work}
  \label{sec:prior-work}

      There are multiple ways to estimate Faraday complexity, including detecting non-linearity in $\chi(\lambda^2)$ \citep{goldstein84faraday}, change in fractional polarisation as a function of frequency \citep{farnes14broadband}, non-sinusoidal variation in fractional polarisation in Stokes $Q$ and $U$ \citep{osullivan12agn}, counting components in the FDF \citep{law11faraday}, minimising the Bayesian information criterion (BIC) over a range of simple and complex models \citep[called `QU fitting';][]{osullivan_broad-band_2017}, the method of Faraday moments \citep{anderson_broadband_2015,Brown11report}, and deep convolutional neural network classifiers \citep[CNNs;][]{brown_classifying_2018}. See \citet{sun15comparison} for a comparison of these methods.

      The most common approaches to estimating complexity are QU fitting \citep[e.g.][]{osullivan_broad-band_2017} and Faraday moments \citep[e.g.][]{anderson_broadband_2015}. To our knowledge there is currently no literature examining the accuracy of QU fitting when applied to complexity classification specifically, though \citet{miyashita19qu} analyse its effectiveness on identifying the structure of two-component sources. \citet{Brown11report} suggested Faraday moments as a method to identify complexity, a method later used by \citet{farnes14broadband} and \citet{anderson_broadband_2015}, but again no literature examines the accuracy. CNNs are the current state-of-the-art with an accuracy of 94.9 per cent \citep{brown_classifying_2018} on simulated ASKAP Band 1 and 3 data, and we will compare our results to this method.

  \subsection{Assumptions on Faraday dispersion functions}
  \label{sec:fdfs}

    \begin{figure}
      \includegraphics[width=\linewidth]{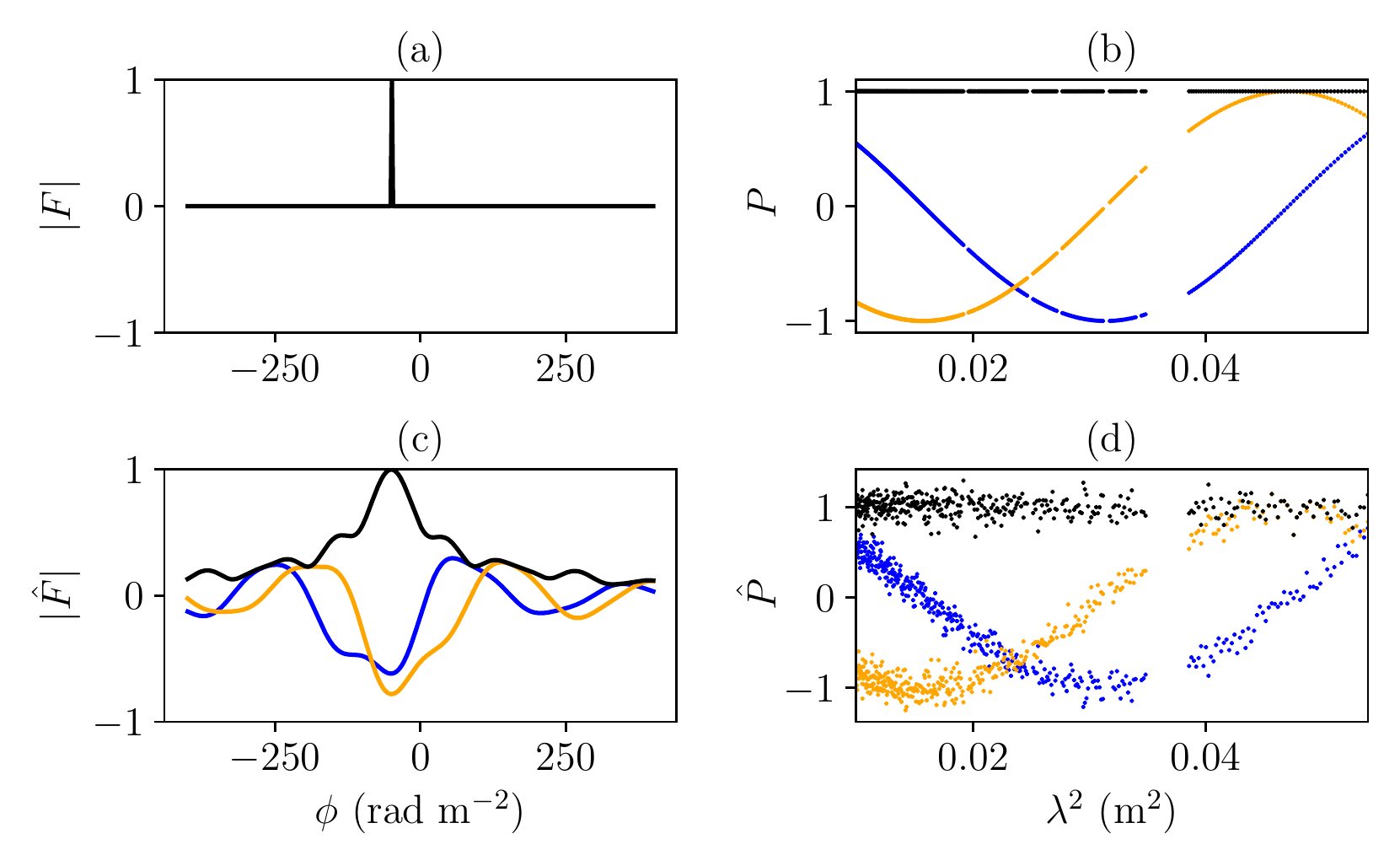}
      \caption{A simple FDF and its corresponding polarised spectra: (a) groundtruth FDF $F$, (b) noise-free polarised spectrum $P$, (c) noisy observed FDF $\hat F$, (d) noisy polarised spectrum $\hat P$. Blue and orange mark real and imaginary components respectively.}
      \label{fig:simple-fdf}
    \end{figure}

    \begin{figure}
      \includegraphics[width=\linewidth]{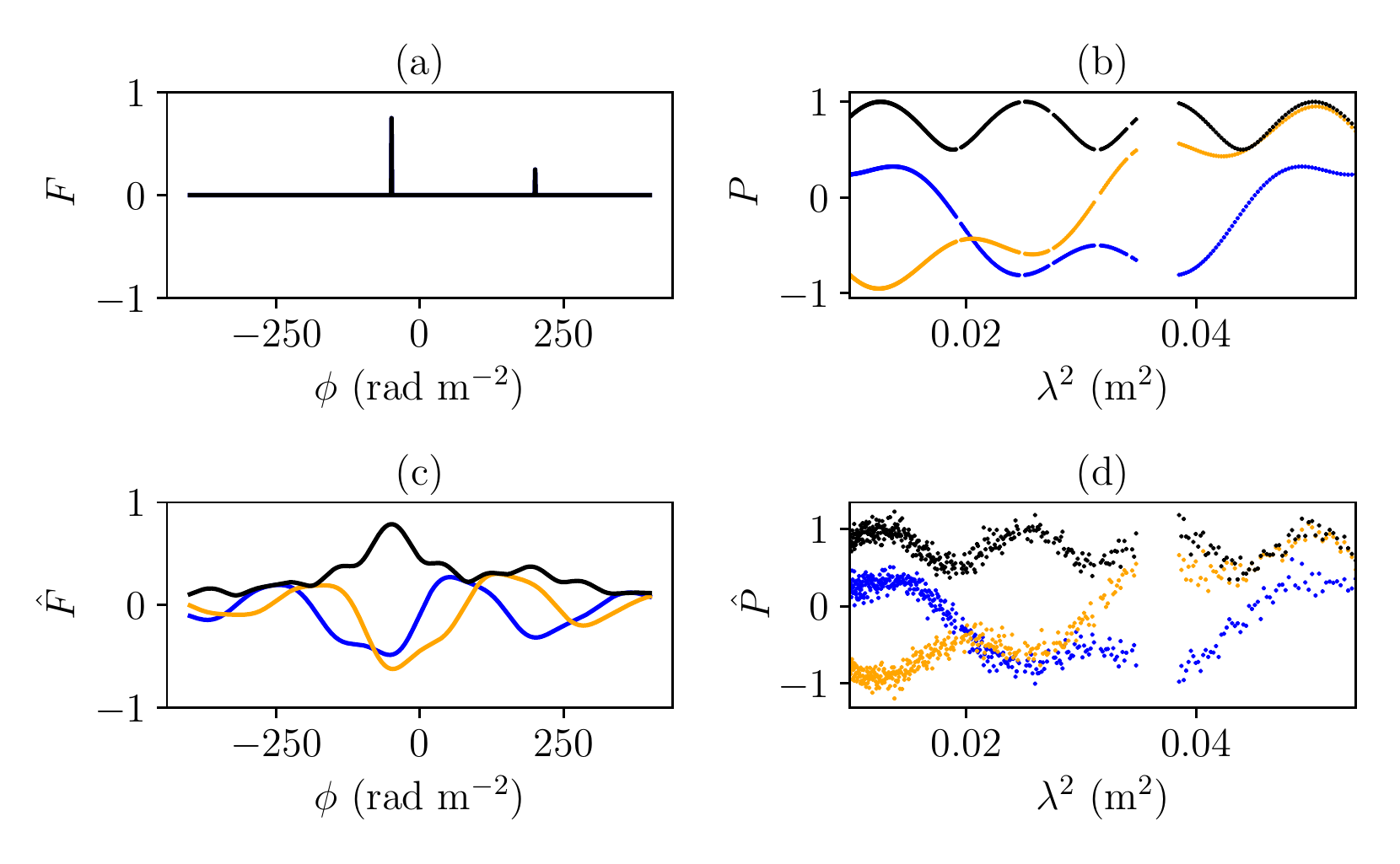}
      \caption{A complex FDF and its corresponding polarised spectra: (a) groundtruth FDF $F$, (b) noise-free polarised spectrum $P$, (c) noisy observed FDF $\hat F$, (d) noisy polarised spectrum $\hat P$. Blue and orange mark real and imaginary components respectively.}
      \label{fig:complex-fdf}
    \end{figure}

    Before we can classify FDFs as Faraday complex or Faraday simple, we need to define FDFs and any assumptions we make about them. An FDF is a function that maps Faraday depth $\phi$ to complex polarisation. It is the distribution of Faraday depths in an observed polarisation spectrum. For a given observation, we assume there is a true, noise-free FDF $F$ composed of at most two Faraday screens. This accounts for most actual sources \citep{anderson_broadband_2015} and extension to three screens would cover most of the remainder---\citet{osullivan_broad-band_2017} found that 89 per cent of their sources were best explained by two or less screens, while the remainder were best explained by three screens. We model the screens by Dirac delta distributions:
    \begin{equation}
        \label{eq:true-fdf}
        F(\phi) = A_0 \delta(\phi - \phi_0) + A_1 \delta(\phi - \phi_1).
    \end{equation}
    $A_0$ and $A_1$ are the polarised flux of each Faraday screen, and $\phi_0$ and $\phi_1$ are the Faraday depths of the respective screens. With this model, a Faraday simple source is one which has $A_0 = 0$, $A_1 = 0$, or $\phi_0 = \phi_1$. By using delta distributions to model each screen, we are assuming that there is no internal Faraday dispersion (which is typically associated with diffuse emission rather than the mostly-compact sources we expect to find in wide-area polarised surveys). $F$ generates a polarised spectrum of the form shown in \autoref{eq:true-pol}:
    \begin{equation}
        \label{eq:true-pol}
        P(\lambda^2) = A_0 e^{2i\phi_0\lambda^2} + A_1 e^{2i\phi_1\lambda^2}.
    \end{equation}
    Such a spectrum would be observed as noisy samples from a number of squared wavelengths $\lambda^2_j, j \in [1, \dots, D]$. We model this noise as a complex Gaussian with standard deviation $\sigma$ and call the noisy observed spectrum $\hat P$:
    \begin{equation}
      \label{eq:noisy-pol}
      \hat P(\lambda_j^2) \sim \mathcal N(P(\lambda^2_j), \sigma^2).
    \end{equation}
    The constant variance of the noise is a simplifying assumption which may not hold for real data, and exploring this is a topic for future work. By performing RM synthesis \citep{brentjens_faraday_2005} on $\hat P$ with uniform weighting we arrive at an observed FDF:
    \begin{equation}
      \label{eq:rm-synthesis}
      \hat F(\phi) = \frac{1}{D} \sum_{j = 1}^D \hat P(\lambda^2_j) e^{-2i\phi\lambda^2_j}.
    \end{equation}
    Examples of $F$, $\hat F$, $P$, and $\hat P$ for simple and complex observations are shown in \autoref{fig:simple-fdf} and \autoref{fig:complex-fdf} respectively. Note that there are two reasons that the observed FDF $\hat F$ does not match the groundtruth FDF $F$. The first is the noise in $\hat P$. The second arises from the incomplete sampling of $\hat P$.

    We do not consider external or internal Faraday dispersion in this work. External Faraday dispersion would broaden the delta functions of \autoref{eq:true-fdf} into peaks, and internal Faraday dispersion would broaden them into top-hat functions. All sources have at least a small amount of dispersion as the Faraday depth is a bulk property of the intervening medium and is subject to noise, but the assumption we make is that this dispersion is sufficiently small that the groundtruth FDFs are well-modelled with delta functions. Faraday thick sources would also invalidate our assumptions, and we assume that there are none in our data as Faraday thickness can be consistent with a two-component model depending on the wavelength sampling \citep[e.g.][]{ma_broad-band_2019,brentjens_faraday_2005}. Nevertheless some external Faraday dispersion would be covered by our model, as depending on observing parameters Faraday thick sources may appear as two screens \citep{vaneck17faraday}.

    To simulate observed FDFs we follow the method of \citet{brown_classifying_2018}, which we describe in \autoref{sec:simulating}.

\section{Classification approach}
\label{sec:approach}

  The Faraday complexity classification problem is as follows: Given an FDF $\hat F$, is it Faraday complex or Faraday simple? In this section we describe the features that we have developed to address this problem, which can be used in any standard machine learning classifier. We trained two classifiers on these features, which we describe here also.

  \subsection{Features}
  \label{sec:scores-method}

    Our features are based on a simple idea: all simple FDFs look essentially the same, up to scaling and translation, while complex FDFs may deviate. A noise-free peak-normalised simple FDF $\hat F_{\mathrm{simple}}$ has the form
    \begin{align}
        \label{eq:f-simple}
        \hat F_{\mathrm{simple}}(\phi; \phi_s) &= R(\phi - \phi_s).
    \end{align}
    where $R$ is the rotation measure spread function (RMSF), the Fourier transform of the wavelength sampling function which is 1 at all observed wavelengths and 0 otherwise. $\phi_s$ traces out a curve in the space of all possible FDFs. In other words, $\hat F_{\mathrm{simple}}$ is a manifold parametrised by $\phi_s$. Our features are derived from relating an observed FDF to the manifold of simple FDFs (the `simple manifold'). We measure the distance of an observed FDF to the simple manifold using distance measure $D_f$, that take all values of the FDF into account:
    \begin{equation}
        \label{eq:complexity-model}
        \varsigma_f(\hat F) = \min_{\phi_s \in \mathbb{R}} D_f\infdivx{\hat F(\phi)}{\hat F_{\mathrm{simple}}(\phi; \phi_s)}.
    \end{equation}
    We propose two distances that have nice properties:
    \begin{itemize}
        \item invariant over changes in complex phase,
        \item translationally invariant in Faraday depth,
        \item zero for Faraday simple sources (i.e. when $A_0 = 0$, $A_1 = 0$, or $\phi_0 = \phi_1$) when there is no noise,
        \item symmetric in components (i.e. swapping $A_0 \leftrightarrow A_1$ and $\phi_0 \leftrightarrow \phi_1$ should not change the distance),
        \item increasing as $A_0$ and $A_1$ become closer to each other, and
        \item increasing as screen separation $|\phi_0 - \phi_1|$ increases over a large range.
    \end{itemize}
    Our features are constructed from this distance and its minimiser. In other words
    we look for the simple FDF $\hat{F}_{\mathrm{simple}}$ that is ``closest'' to the observed FDF $\hat{F}$.
    The minimiser $\phi_s$ is the Faraday depth of the simple FDF.

    While we could choose any distance that operates on functions, we used the 2-Wasserstein ($W_2$) distance \eqref{eq:W2-distance} and the Euclidean distance \eqref{eq:Euclidean-distance}. The $W_2$ distance operates on probability distributions and can be thought of as the minimum cost to `move' one probability distribution to the other, where the cost of moving one unit of probability mass is the squared distance it is moved. Under $W_2$ distance, the minimiser $\phi_w$ in \autoref{eq:complexity-model} can be interpreted as the Faraday depth that the FDF $\hat F$ would be observed to have if its complexity was unresolved (i.e. the weighted mean of its components). The Euclidean distance is the square root of the least-squares loss which is often used for fitting $\hat{F}_{\mathrm{simple}}$ to the FDF $\hat F$. Under Euclidean distance, the minimiser $\phi_s$ is equivalent to the depth of the best-fitting single component under assumption of Gaussian noise in $\hat F$.
    We calculated the $W_2$ distance using \texttt{Python Optimal Transport} \citep{flamary17pot}, and we calculated the Euclidean distance using \texttt{scipy.spatial.distance.euclidean} \citep{scipy2020}.
    Further intuition about the two distances is provided in \autoref{sec:interpreting-distances}.

\begin{figure}
      \centering
      \includegraphics[width=\linewidth]{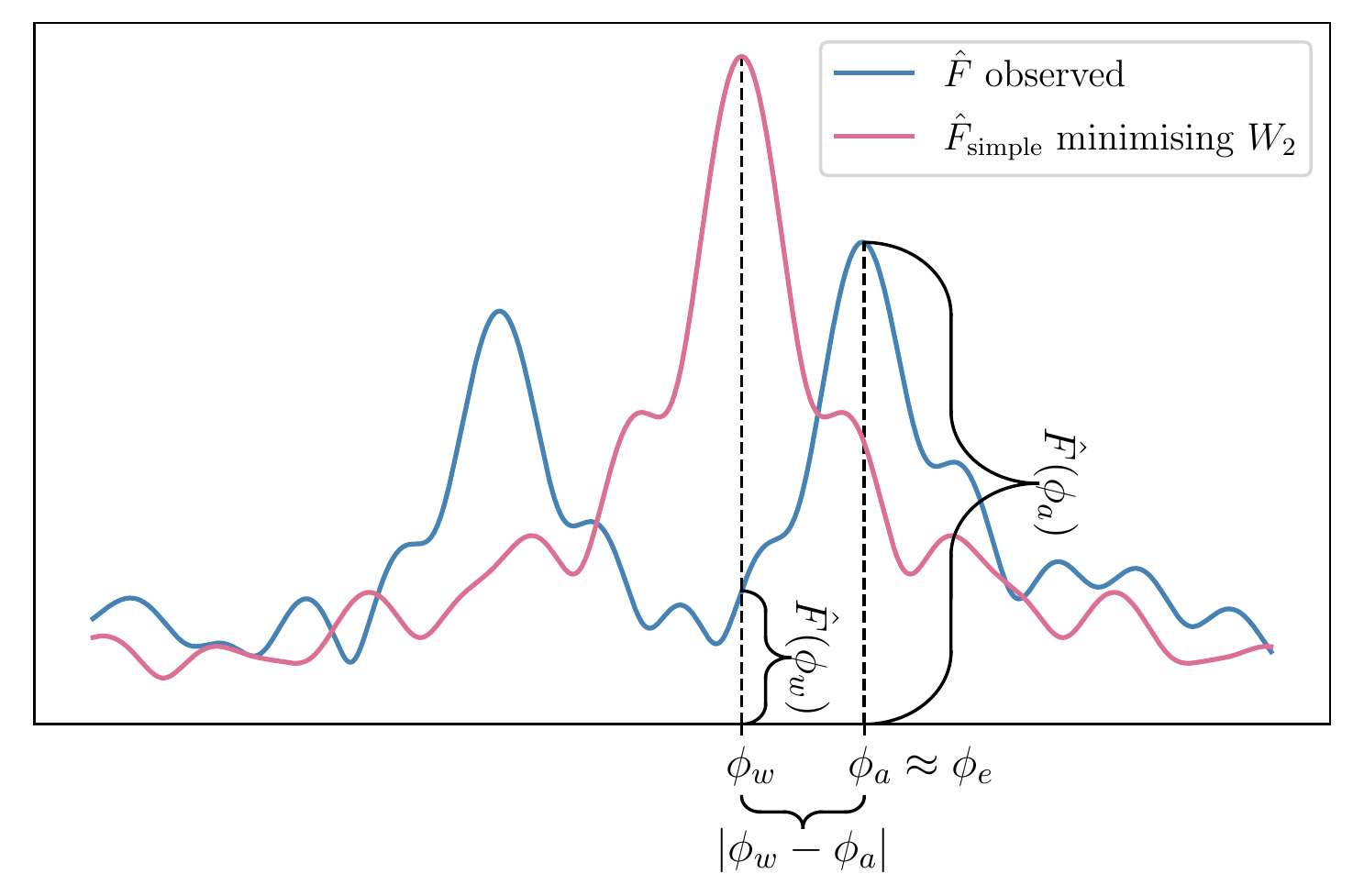}
      \caption{\label{fig:features-on-fdf} An example of how an observed FDF $\hat F$ relates to our features. $\phi_w$ is the $W_2$-minimising Faraday depth, and $\phi_a$ is the $\hat F$-maximising Faraday depth (approximately equal to the Euclidean-minimising Faraday depth). The remaining two features are the $W_2$ and Euclidean distances between the depicted FDFs.}
    \end{figure}

    We denote by $\phi_w$ and $\phi_e$, the Faraday depth of the simple FDF that minimises the respective distances
    (2-Wasserstein and Euclidean).
    \begin{align*}
       \phi_w &= \underset{\phi_w}{\mathrm{argmin}}\ D_{W_2}\infdivx{\hat F(\phi)}{\hat F_{\mathrm{simple}}(\phi; \phi_w)},\\
       \phi_e &= \underset{\phi_e}{\mathrm{argmin}}\ D_E\infdivx{\hat F(\phi)}{\hat F_{\mathrm{simple}}(\phi; \phi_e)}.
     \end{align*}
     These features are depicted on an example FDF in \autoref{fig:features-on-fdf}.
     For simple observed FDFs, the fitted Faraday depths $\phi_w$ and $\phi_e$ both tend to be
     close to the peak of the observed FDF. However for complex observed FDFs, $\phi_w$ tends
     to be at the average depth between the two major peaks of the observed FDF, being closer
     to the higher peak. For notation convenience, we denote the Faraday depth of the
     observed FDF that has largest magnitude as $\phi_a$, i.e.
     \begin{equation*}
       \phi_a = \underset{\phi_a}{\mathrm{argmax}}\ |\hat F(\phi_a)|,\\
     \end{equation*}
     Note that in practice $\phi_a \approx \phi_e$.
     For complex observed FDFs, the values of Faraday depths $\phi_w$ and $\phi_a$ tend
     to differ (essentially by a proportion of the location of the second screen).
     The difference between $\phi_w$ and $\phi_a$ therefore provides useful information
     to identify complex FDFs.
     When the observed FDF is simple, the 2-Wasserstein fit will overlap significantly,
     hence the observed magnitudes $\hat F(\phi_w)$ and $\hat F(\phi_a)$ will be similar.
     However, for complex FDFs $\phi_w$ and $\phi_a$ are at different depths,
     leading to different values of $\hat F(\phi_w)$ and $\hat F(\phi_a)$.
     Therefore the magnitudes of the observed FDFs at the depths $\phi_w$ and $\phi_a$
     indicate how different the observed FDF is from a simple FDF.

    In summary, we provide the following features to the classifier:
    \begin{itemize}
      \item $\log |\phi_w - \phi_a|$,
      \item $\log \hat F(\phi_w)$,
      \item $\log \hat F(\phi_a)$,
      \item $\log D_{W_2}\infdivx{\hat F(\phi)}{\hat F_{\mathrm{simple}}(\phi; \phi_w)}$,
      \item $\log D_{E}\infdivx{\hat F(\phi)}{\hat F_{\mathrm{simple}}(\phi; \phi_e)}$,
    \end{itemize}
    where $D_E$ is the Euclidean distance, $D_{W_2}$ is the $W_2$ distance, $\phi_a$ is the Faraday depth of the FDF peak, $\phi_w$ is the minimiser for $W_2$ distance, and $\phi_e$ is the minimiser for Euclidean distance.

    \subsection{Interpreting distances}
    \label{sec:interpreting-distances}

    Interestingly, in the case where there is no RMSF, \autoref{eq:complexity-model} with $W_2$ distance reduces to the Faraday moment already in common use:
    \begin{align}
        D_{W_2}(F) &= \min_{\phi_w \in \mathbb{R}} D_{W_2}\infdivx{F(\phi)}{F_{\mathrm{simple}}(\phi; \phi_w)}\label{eq:W2-distance}\\
            &= \left(\frac{A_0A_1}{(A_0 + A_1)^2} (\phi_0 - \phi_1)^2\right)^{1/2}.
    \end{align}
    See \autoref{sec:w2-to-faraday-moments} for the corresponding calculation. In this sense, the $W_2$ distance can be thought of as a generalised Faraday moment, and conversely an interpretation of Faraday moments as a distance from the simple manifold in the case where there is no RMSF. Euclidean distance behaves quite differently in this case, and the resulting distance measure is totally independent of Faraday depth:
    \begin{align}
        D_{E}(F) &= \min_{\phi_e \in \mathbb{R}} D_E\infdivx{F(\phi)}{F_{\mathrm{simple}}(\phi; \phi_e)}\label{eq:Euclidean-distance}\\
            &= \sqrt{2} \frac{\min(A_0, A_1)}{A_0 + A_1}.
    \end{align}
    See \autoref{sec:euclidean-calculation} for the corresponding calculation.

  \subsection{Classifiers}
  \label{sec:classifiers}

    We trained two classifiers on simulated observations using these features: logistic regression (LR) and extreme gradient boosted trees (XGB). These classifiers are useful together for understanding Faraday complexity classification. LR is a linear classifier that is readily interpretable by examining the weights it applies to each feature, and is one of the simplest possible classifiers. XGB is a powerful off-the-shelf non-linear ensemble classifier, and is an example of a decision tree ensemble which are widely used in astronomy \citep[e.g.][]{valle20shap,hlozek20lsst}. We used the \texttt{scikit-learn} implementation of LR and we use the \texttt{XGBoost} library for XGB. We optimised hyperparameters for XGB using a fork of \texttt{xgboost-tuner} \footnote{\url{https://github.com/chengsoonong/xgboost-tuner}} as utilised by \citet{zhu20mutagenic}. We used 1~000 iterations of randomised parameter tuning and the hyperparameters we found are tabulated in \autoref{tab:hyperparameters-xgb}. We optimised hyperparameters for LR using a 5-fold cross-validation grid search implemented in \texttt{sklearn.model\textunderscore{}selection.GridSearchCV}. The resulting hyperparameters are tabulated in \autoref{tab:hyperparameters-lr} in the Appendix.

\section{Experimental method and results}
\label{sec:experiment-classification}

  We applied our classifiers to classify simulated (\autoref{sec:cnn-comparison} and \ref{sec:results-simulated}) and real (\autoref{sec:results-observed}) FDFs. We replicated the experimental setup of \citet{brown_classifying_2018} for comparison with the state-of-the-art CNN classification method, and we also applied our method to 142 real FDFs observed with the Australia Telescope Compact Array (ATCA) from Livingston et al. (2020, submitted) and \citet{osullivan_broad-band_2017}.

  \subsection{Data}

  \subsubsection{Simulated training and validation data}
  \label{sec:simulated-training-data}

    Our classifiers were trained and validated on simulated FDFs. We produced two sets of simulated FDFs, one for comparison with the state-of-the-art method in the literature and one for application to our observed FDFs (described in \autoref{sec:observational-data}). We refer to the former as the `ASKAP' dataset as it uses frequencies from the Australian Square Kilometre Array Pathfinder 12-antenna early science configuration. These frequencies included 900 channels from 700--1300 and 1500--1800~MHz and were used to generate simulated training and validation data by \citet{brown_classifying_2018}. We refer to the latter as the `ATCA' dataset as it uses frequencies from the 1--3~GHz configuration of the ATCA. These frequencies included 394 channels from 1.29--3.02~GHz and match our real data. We simulated Faraday depths from $-50$ to $50$ rad~m$^{-2}$ for the `ASKAP' dataset (matching Brown) and $-500$ to $500$ for the `ATCA' dataset.

    For each dataset, we simulated 100~000 FDFs, approximately half simple and half complex. We randomly allocated half of these FDFs to a training set and reserved the remaining half for validation. Each FDF had complex Gaussian noise added to the corresponding polarisation spectrum. For the `ASKAP' dataset, we sampled the standard deviation of the noise uniformly between 0 and $\sigma_{\max} = 0.333$, matching the dataset of \citet{brown_classifying_2018}.
    For the `ATCA' dataset, we fit a log-normal distribution to the standard deviations of O'Sullivan's data \citep{osullivan_broad-band_2017} from which we sampled our values of $\sigma$:
    \begin{equation}
      \sigma \sim \frac{1}{0.63 \sqrt{2 \pi} \sigma} \exp \left(-\frac{\log\left(50 \sigma - 0.5\right)^2}{2 \times 0.63^2}\right)
    \end{equation}

  \subsubsection{Observational data}
  \label{sec:observational-data}

    We used two real datasets containing a total of 142 sources: 42 polarised spectra from Livingston et al. (2020, submitted) and 100 polarised spectra from \citet{osullivan_broad-band_2017}. These datasets were observed in similar frequency ranges on the same telescope (with different binning), but are in different parts of the sky. The Livingston data were taken near the Galactic Centre, and the O'Sullivan data were taken away from the plane of the Galaxy. There are more Faraday complex sources near the Galactic Centre compared to more Faraday simple sources away from the plane of the Galaxy (Livingston et al.). The similar frequency channels used in the two datasets result in almost identical RMSFs over the Faraday depth range we considered (-500 to 500 rad m$^{-2}$), so we expected that the classifiers would work equally well on both datasets with no need to re-train. We discarded the 26 Livingston sources with modelled Faraday depths outside of this Faraday depth range, which we do not expect to affect the applicability of our methods to wide-area surveys because these fairly high depths are not common.

    \citet{livingston21faraday} used RM-CLEAN \citep{heald09faraday} to identify significant components in their FDFs. Some of these components had very high Faraday depths up to 2000 rad m$^{-2}$, but we chose to ignore these components in this paper as they are much larger than might be expected in a wide-area survey like POSSUM. They used the second Faraday moment \citep{Brown11report} to estimate Faraday complexity, with Faraday depths determined using \texttt{scipy.signal.find\textunderscore{}peaks} on the cleaned FDFs, with a cutoff of 7 times the noise of the polarised spectrum. Using this method, they estimated that 89 per cent of their sources were Faraday complex i.e. had a Faraday moment greater than 0.

    \citet{osullivan_broad-band_2017} used the QU-fitting and model selection technique described in \citet{osullivan12agn}. The QU-fitting models contained up to three Faraday screen components as well as a term for internal and external Faraday dispersion. We ignore the Faraday thickness and dispersion for the purposes of this paper, as most sources were not found to have Faraday thickness and dispersion is beyond the scope of our current work. 37 sources had just one component, 52 had two, and the remaining 11 had three.

  \subsection{Results on `ASKAP' dataset}
  \label{sec:cnn-comparison}

    \begin{table}[htbp]
      \small
      \caption{\label{tab:cms} Confusion matrix entries for LR and XGB on `ASKAP' and `ATCA' simulated datasets, and the CNN confusion matrix entries adapted from \citet{brown_classifying_2018}.}
      \begin{tabular}{r|ccc|cc}
        \hline\hline
        & \multicolumn{3}{c}{`ASKAP'} & \multicolumn{2}{c}{`ATCA'} \\
        & LR & XGB & CNN & LR & XGB \\
        True negative rate & 0.99 & 0.99 & 0.97 & 0.92 & 0.91\\
        False positive rate & 0.01 & 0.01 & 0.03 & 0.08 & 0.09\\
        False negative rate & 0.10 & 0.09 & 0.07 & 0.16 & 0.10\\
        True positive rate & 0.90 & 0.91 & 0.93 & 0.84 & 0.90\\
        \hline\hline
      \end{tabular}\\
    \end{table}

    The accuracy of the LR and XGB classifiers on the `ASKAP' testing set was 94.4 and 95.1 per cent respectively. The rates of true and false identifications are summarised in \autoref{tab:cms}. These results are very close to the CNN presented by \citet{brown_classifying_2018}, with a slightly higher true negative rate and a slightly lower true positive rate (recalling that positive sources are complex, and negative sources are simple). The accuracy of the CNN was 94.9, slightly lower than our XGB classifier and slightly higher than our LR classifier. Both of our classifiers therefore produce similar classification performance to the CNN, with faster training time and easier interpretation.

  \subsection{Results on `ATCA' dataset}
  \label{sec:results-simulated}

    The accuracy of the LR and XGB classifiers on the `ATCA' dataset was 89.2 and 90.5 per cent respectively. The major differences between the `ATCA' and the `ASKAP' experiments are the range of the simulated Faraday depths and the distribution of noise levels. The `ASKAP' dataset, to match past CNN work, only included depths from $-50$ to $50$ rad m$^{-2}$, while the `ATCA' dataset includes depths from $-500$ to $500$ rad m$^{-2}$. The rates of true and false identifications are again shown in \autoref{tab:cms}.

    \begin{figure}
      \begin{subfigure}{\linewidth}
        \includegraphics[width=\linewidth]{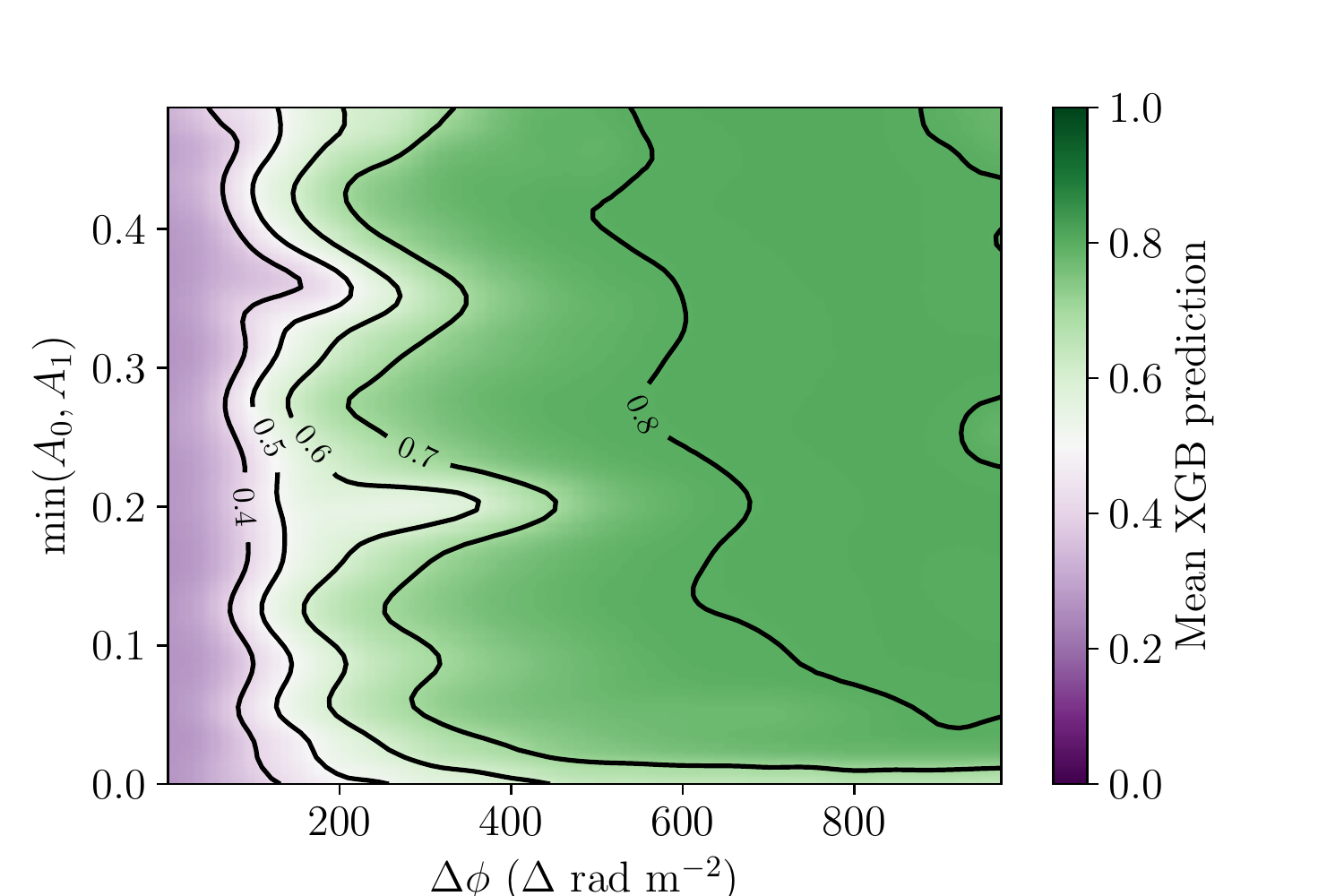}
        \caption{\label{fig:mean-xgb-pred}}
      \end{subfigure}
      \begin{subfigure}{\linewidth}
        \includegraphics[width=\linewidth]{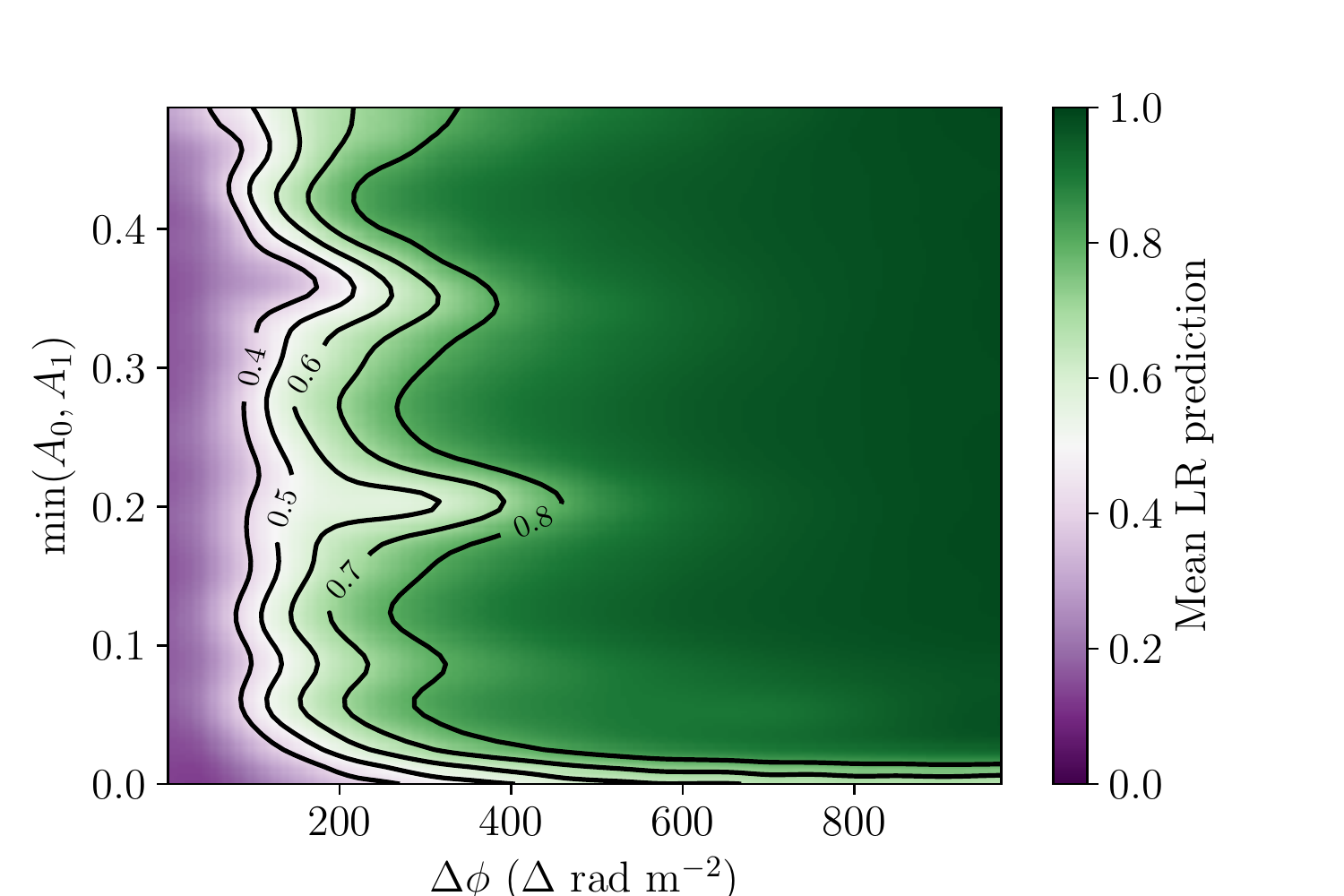}
        \caption{\label{fig:mean-lr-pred}}
      \end{subfigure}
      \caption{\label{fig:amps-dphi-mean} Mean prediction as a function of component depth separation and minimum component amplitude for (a) XGB and (b) LR.}
    \end{figure}

    As we know the true Faraday depths of the components in our simulation, we can investigate the behaviour of these classifiers as a function of physical properties. \autoref{fig:amps-dphi-mean} shows the mean classifier prediction as a function of component depth separation and minimum component amplitude. This is tightly related to the mean accuracy, as the entire plot domain contains complex spectra besides the left and bottom edge: by thresholding the classifier prediction to a certain value, the accuracy will be one hundred per cent on the non-edge for all sources with higher prediction values.

  \subsection{Results on observed FDFs}
  \label{sec:results-observed}

    \begin{figure*}
      \centering
      \includegraphics[width=\linewidth]{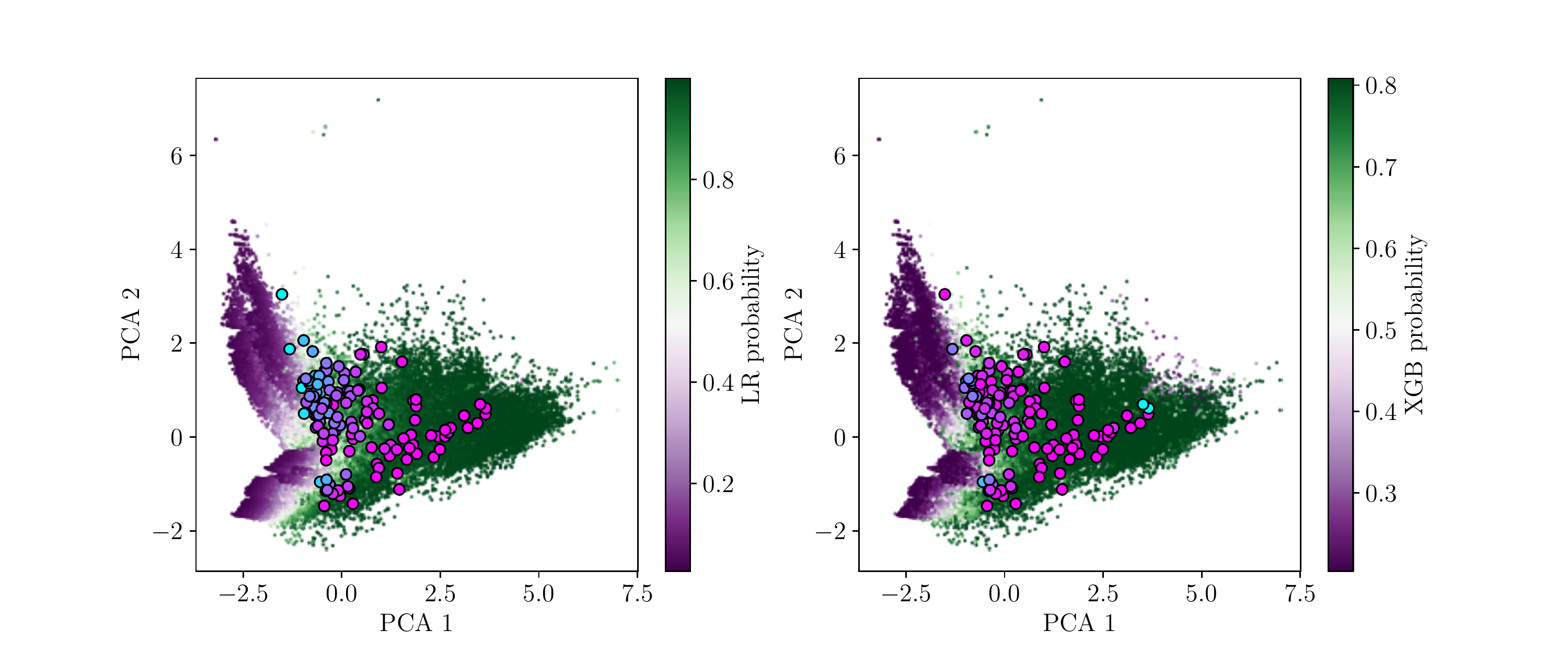}
      \caption{Principal component analysis for simulated data (coloured dots) with observations overlaid (black-edged circles). Observations are coloured by their XGB or LR estimated probability of being complex, with blue indicating `most simple' and pink indicating `most complex'.}
      \label{fig:pca}
    \end{figure*}
    \begin{figure}
      \centering
      \includegraphics[width=\linewidth]{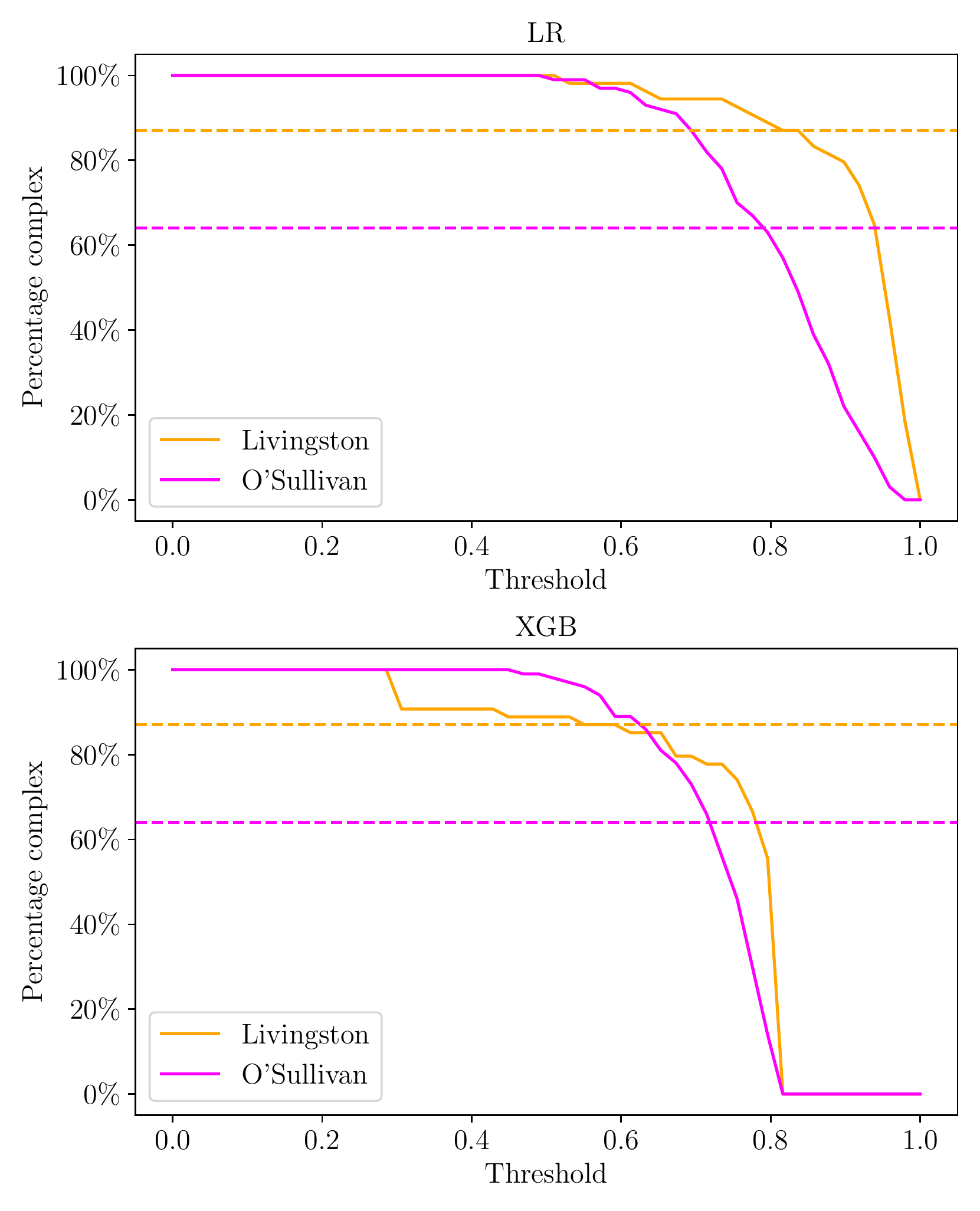}
      \caption{Estimated rates of Faraday complexity for the Livingston and O'Sullivan datasets as functions of threshold. The horizontal lines indicate the rates of Faraday complexity estimated by Livingston and O'Sullivan respectively.}
      \label{fig:complexity-rates}
    \end{figure}

    We used the LR and XGB classifiers which were trained on the `ATCA' dataset to estimate the probability that our 142 observed FDFs (\autoref{sec:observational-data}) were Faraday complex. As these classifiers were trained on simulated data, they face the issue of the `domain gap': the distribution of samples from a simulation differs from the distribution of real sources, and this affects performance on real data. Solving this issue is called `domain adaptation' and how to do this is an open research question in machine learning \citep{zhang2019transfer,pan10transfer}. Nevertheless, the features of our observations mostly fall in the same region of feature space as the simulations (\autoref{fig:pca}) and so we expect reasonably good domain transfer.

    Two apparently complex sources in the Livingston sample are classified as simple with high probability by XGB. These outliers are on the very edge of the training sample (\autoref{fig:pca}) and the underdensity of training data here is likely the cause of this issue. LR does not suffer the same issue, producing plausible predictions for the entire dataset, and these sources are instead classified as complex with high probability.

    With a threshold of 0.5, LR predicted that 96 and 83 per cent of the Livingston and O'Sullivan sources were complex respectively. This is in line with expectations that the Livingston data should have more Faraday complex sources than the O'Sullivan data due to their location near the Galactic Centre. XGB predicted that 93 and 100 per cent of the Livingston and O'Sullivan sources were complex respectively. \citet{livingston21faraday} found that 90 per cent of their sources were complex, and \citet{osullivan_broad-band_2017} found that 64 per cent of their sources were complex. This suggests that our classifiers are overestimating complexity, though it could also be the case that the methods used by Livingston and O'Sullivan underestimate complexity. Modifying the prediction threshold from 0.5 changes the estimated rate of Faraday complexity, and we show the estimated rates against threshold for both classifiers in \autoref{fig:complexity-rates}. We suggest that this result is indicative of our probabilities being uncalibrated, and a higher threshold should be chosen in practice. We chose to keep the threshold at 0.5 as this had the highest accuracy on the simulated validation data. The very high complexity rates of XGB and two outlying classifications indicate that the XGB classifier may be overfitting to the simulation and that it is unable to generalise across the domain gap.

    \autoref{fig:all-observed-fdfs-lr} and \autoref{fig:all-observed-fdfs-xgb} show every observed FDF ordered by estimated Faraday complexity, alongside the models predicted by Livingston and \citet{osullivan_broad-band_2017}, for LR and XGB respectively. There is a clear visual trend of increasingly complex sources with increasing predicted probability of being complex.

\section{Discussion}
\label{sec:discussion}

  On simulated data (\autoref{sec:results-simulated}) we achieve state-of-the-art accuracy. Our results on observed FDFs show that our classifiers produce plausible results, with \autoref{fig:all-observed-fdfs-lr} and \autoref{fig:all-observed-fdfs-xgb} showing a clear trend of apparent complexity. Some issues remain: we discuss the intrinsic overlap between simple and complex FDFs in \autoref{sec:overlap} and the limitations of our method in \autoref{sec:limitations}.

  \subsection{Complexity and seeming `not simple'}
  \label{sec:overlap}

    Through this work we found our methods limited by the significant overlap between complex and simple FDFs. Complex FDFs can be consistent with simple FDFs due to close Faraday components or very small amplitudes on the secondary component, and vice versa due to noise.

    The main failure mode of our classifiers is misclassifying a complex source as simple (\autoref{tab:cms}). Whether sources with close components or small amplitudes should be considered complex is not clear, since for practical purposes they can be treated as simple: assuming the source is simple yields a very similar RM to the RM of the primary component, and thus would not negatively impact further data products such as an RM grid. The scenarios where we would want a Faraday complexity classifier rather than a polarisation structure model -- large-scale analysis and wide-area surveys -- do not seem to be disadvantaged by considering such sources simple. Additional sources similar to these are likely hidden in presumably `simple' FDFs by the frequency range and spacing of the observations, just as how these complex sources would be hidden in lower-resolution observations. Note also that misidentification of complex sources as simple is intrinsically a problem with complexity estimation even for models not well-represented by a simple FDF, as complex sources may conspire to appear as a wide range of viable models including simple \citep{sun15comparison}.

    Conversely, high-noise simple FDFs may be consistent with complex FDFs. One key question is how Faraday complexity estimators should behave as the noise increases: should high noise result in a complex prediction or a simple prediction, given that a complex or simple FDF would both be consistent with a noisy FDF? Occam's razor suggests that we should choose the simplest suitable model, and so increasing noise should lead to predictions of less complexity. This is not how our classifiers operate, however: high-noise FDFs are different to the model simple FDFs and so are predicted to be `not simple'. In some sense our classifiers are not looking for complex sources, but are rather looking for `not simple' sources.

  \subsection{Limitations}
  \label{sec:limitations}

    Our main limitations are our simplifying assumptions on FDFs and the domain gap between simulated and real observations. However, our proposed features (Section~\autoref{sec:scores-method}) can be applied to future improved simulations.

    It is unclear what the effect of our simplifying assumptions are on the effectiveness of our simulation. The three main simplifications that may negatively affect our simulations are 1) limiting to two components, 2) assuming no external Faraday dispersion, and 3) assuming no internal Faraday dispersion (Faraday thickness). Future work will explore removing these simplifying assumptions, but will need to account for the increased difficulty in characterising the simulation with more components and no longer having Faraday screens as components. Additionally, more work will be required to make sure that the rates of internal and external Faraday dispersion match what might be expected from real sources, or risk making a simulation that has too large a range of consistent models for a given source: for example, a two-component source could also be explained as a sufficiently wide or resolved-out Faraday thick source or a three-component source with a small third component. This greatly complicates the classification task.

    Previous machine learning work \citep[e.g.][]{brown_classifying_2018} has not been run before on real FDF data, so this paper is the first example of the domain gap arising in Faraday complexity classification. This is a problem that requires further research to solve. We have no good way to ensure that our simulation matches reality, so some amount of domain adaptation will always be necessary to train classifiers on simulated data and then apply these classifiers to real data. But with the low source counts in polarisation science (high-resolution spectropolarimetric data currently numbers in the few hundreds) any machine learning method will need to be trained on simulations. This is not just a problem in Faraday complexity estimation, and domain adaptation is also an issue faced in the wider astroinformatics community: large quantities of labelled data are hard to come by, and some sources are very rare \citep[e.g. gravitational wave detections or fast radio bursts;][]{zevin17gravityspy, gebhard19convolutional, agarwal20fetch}. LR seems to handle the domain adaptation better than XGB, with only a slightly lower accuracy on simulated data. Our results are plausible and the distribution of our simulation well overlaps the distribution of our real data (\autoref{fig:pca}).

\section{Conclusion}
\label{sec:conclusion}

  We developed a simple, interpretable machine learning method for estimating Faraday complexity. Our interpretable features were derived by comparing observed FDFs to idealised simple FDFs, which we could determine both for simulated and real observations. We demonstrated the effectiveness of our method on both simulated and real data. Using simulated data, we found that our classifiers were 95 per cent accurate, with near perfect recall (specificity) of Faraday simple sources. On simulated data that matched existing observations, our classifiers obtained an accuracy of 90 per cent. Evaluating our classifiers on real data gave the plausible results shown in \autoref{fig:all-observed-fdfs-lr}, and marks the first application of machine learning to observed FDFs. Future work will need to narrow the domain gap to improve transfer of classifiers trained on simulations to real, observed data.

\begin{acknowledgements}
    This research was conducted in Canberra, on land for which the Ngunnawal and Ngambri people are the traditional and ongoing custodians. M.J.A. and J.D.L. were supported by the Australian Government Research Training Program. M.J.A. was supported by the Astronomical Society of Australia. The Australia Telescope Compact Array is part of the Australia Telescope National Facility which is funded by the Australian Government for operation as a National Facility managed by CSIRO. We acknowledge the Gomeroi people as the traditional owners of the Observatory site. We thank the anonymous referee for their comments on this work.
\end{acknowledgements}

\newpage
\begin{appendix}
  \section{2-Wasserstein begets Faraday moments}
  \label{sec:w2-to-faraday-moments}
    Minimising the 2-Wasserstein distance between a model FDF and the simple manifold gives the second Faraday moment of that FDF. Let $\tilde F$ be the sum-normalised model FDF and let $\tilde S$ be the sum-normalised simple model FDF:
    \begin{align}
      \tilde F(\phi) &= \frac{A_0 \delta(\phi - \phi_0) + A_1 \delta(\phi - \phi_1)}{A_0 + A_1}\\
      \tilde S(\phi; \phi_w) &= \delta(\phi - \phi_w).
    \end{align}
    The $W_2$ distance, usually defined on probability distributions, can be extended to one-dimensional complex functions $A$ and $B$ by normalising them:
      \begin{align}
        \label{eq:w2}
        D_{W_2}\infdivx{A}{B}^2 &= \inf_{\gamma \in \Gamma(A, B)} \iint_{\phi_{\min}}^{\phi_{\max}} |x - y|^2\ \mathrm{d}\gamma(x, y) \\
        \label{eq:normalised}
        \tilde A(\phi) &= \frac{|A(\phi)|}{\int_{\phi_{\min}}^{\phi_{\max}} |A(\theta)|\ \mathrm{d}\theta}\\
        \tilde B(\phi) &= \frac{|B(\phi)|}{\int_{\phi_{\min}}^{\phi_{\max}} |B(\theta)|\ \mathrm{d}\theta}
      \end{align}
      where $\Gamma(A, B)$ is the set of couplings of $A$ and $B$, i.e. the set of joint probability distributions that marginalise to $A$ and $B$; and $\inf_{\gamma \in \Gamma(A, B)}$ is the infimum over $\Gamma(A, B)$. This can be interpreted as the minimum cost to `move' one probability distribution to the other, where the cost of moving one unit of probability mass is the squared distance it is moved.

    The set of couplings $\Gamma(\tilde F, \tilde S)$ is the set of all joint probability distributions $\gamma$ such that
    \begin{align}
      \int_{\phi_{\min}}^{\phi_{\max}} \gamma(\phi, \varphi)\ \mathrm{d}\phi &= \tilde S(\varphi; \phi_w),\\
      \int_{\phi_{\min}}^{\phi_{\max}} \gamma(\phi, \varphi)\ \mathrm{d}\varphi &= \tilde F(\phi).
    \end{align}
    The coupling that minimises the integral in \autoref{eq:w2} will be the optimal transport plan between $\tilde F$ and $\tilde S$. Since $\tilde F$ and $\tilde S$ are defined in terms of delta functions, the optimal transport problem reduces to a discrete optimal transport problem and the optimal transport plan is:
    \begin{equation}
      \gamma(\phi, \varphi) = \frac{A_0 \delta(\phi - \phi_0) + A_1 \delta(\phi - \phi_1)}{A_0 + A_1} \delta(\varphi - \phi_w).
    \end{equation}
    In other words, to move the probability mass of $\tilde S$ to $\tilde F$, a fraction $A_0/(A_0 + A_1)$ is moved from $\phi_w$ to $\phi_0$ and the complementary fraction $A_1/(A_0 + A_1)$ is moved from $\phi_w$ to $\phi_1$. Then:
    \begin{align}
      D_{W_2}\infdivx{\tilde F}{\tilde S}^2 &= \iint_{\phi_{\min}}^{\phi_{\max}} |\phi - \varphi|^2\ \mathrm{d}\gamma(\phi, \varphi)\\
        &= \frac{A_0 (\phi_0 - \phi_w)^2 + A_1 (\phi_1 - \phi_w)^2}{A_0 + A_1}.
    \end{align}
    To obtain the $W_2$ distance to the simple manifold, we need to minimise this over $\phi_w$. Differentiate with respect to $\phi_w$ and set equal to zero to find
    \begin{equation}
      \phi_w = \frac{A_0 \phi_0 + A_1 \phi_1}{A_0 + A_1}.
    \end{equation}
    Substituting this back in, we find
    \begin{align}
      \varsigma_{W_2}(F)^2 &= \frac{A_0 A_1}{A_0 + A_1}(\phi_0 - \phi_1)^2
    \end{align}
    which is the Faraday moment.

\section{Euclidean distance in the no-RMSF case}
\label{sec:euclidean-calculation}

  In this section we calculate the minimumised Euclidean distance evaluated on a model FDF (\autoref{eq:true-fdf}). Let $\tilde F$ be the sum-normalised model FDF and let $\tilde S$ be the normalised simple model FDF:
  \begin{align}
    \tilde F(\phi) &= \frac{A_0 \delta(\phi - \phi_0) + A_1 \delta(\phi - \phi_1)}{A_0 + A_1}\\
    \tilde S(\phi; \phi_e) &= \delta(\phi - \phi_e).
  \end{align}

  The Euclidean distance between $\tilde F$ and $\tilde S$ is then
  \begin{align}
    &D_E\infdivx{\tilde F(\phi)}{\tilde S(\phi; \phi_e)}^2\\
    &= \int_{\phi_{\min}}^{\phi_{\max}} \left|\tilde F(\phi) - \delta(\phi - \phi_e) \right|^2\ \mathrm{d}\phi.
  \end{align}

  Assume $\phi_0 \neq \phi_1$ (otherwise, $D_E$ will always be either $0$ or $2$). If $\phi_e = \phi_0$, then
  \begin{align}
    &D_E\infdivx{\tilde F(\phi)}{\tilde S(\phi; \phi_e)}^2\\
      &= \frac{1}{(A_0 + A_1)^2} \int_{\phi_{\min}}^{\phi_{\max}} A_1^2 \left|\delta(\phi - \phi_1) - \delta(\phi - \phi_0) \right|^2\ \mathrm{d}\phi\\
      &= \frac{2 A_1^2}{(A_0 + A_1)^2}
  \end{align}
  and similarly for $\phi_e = \phi_1$. If $\phi_e \neq \phi_0$ and $\phi_e \neq \phi_1$, then
  \begin{equation}
    D_E\infdivx{\tilde F(\phi)}{\tilde S(\phi; \phi_e)}^2 = \frac{A_0^2 + A_1^2 + 1}{(A_0 + A_1)^2}.
  \end{equation}
  The minimised Euclidean distance when $\phi_0 \neq \phi_1$ is therefore
  \begin{align}
      D_E(F) &= \min_{\phi_e \in \mathbb{R}} D_E\infdivx{F(\phi)}{F_{\mathrm{simple}}(\phi; \phi_e)}\\
          &= \sqrt{2} \frac{\min(A_0, A_1)}{A_0 + A_1}.
  \end{align}
  If $\phi_0 = \phi_1$, then the minimised Euclidean distance is 0.

\section{Hyperparameters for LR and XGB}
\label{sec:hyperparameters}

  This section contains tables of the hyperparameters that we used for our classifiers. \autoref{tab:hyperparameters-xgb} and \autoref{tab:hyperparameters-lr} tabulate the hyperparameters for XGB and LR respectively for the `ATCA' dataset. \autoref{tab:hyperparameters-xgb-askap12} and \autoref{tab:hyperparameters-lr-askap12} tabulate the hyperparameters for XGB and LR respectively for the `ASKAP' dataset.

  \begin{table}[htbp]
    \caption{\label{tab:hyperparameters-xgb} XGB hyperparameters for the `ATCA' dataset.}
    \begin{tabular}{ll}
      \hline\hline
      Parameter & Value\\\hline
      colsample\textunderscore{}bytree & 0.912\\
      gamma & 0.532\\
      learning\textunderscore{}rate & 0.1\\
      max\textunderscore{}depth & 7\\
      min\textunderscore{}child\textunderscore{}weight & 2\\
      scale\textunderscore{}pos\textunderscore{}weight & 1\\
      subsample & 0.557\\
      n\textunderscore{}estimators & 135\\
      reg\textunderscore{}alpha & 0.968\\
      reg\textunderscore{}lambda & 1.420\\
      \hline\hline
    \end{tabular}
  \end{table}

  \begin{table}[htbp]
    \caption{\label{tab:hyperparameters-lr} LR hyperparameters for the `ATCA' dataset.}
    \begin{tabular}{ll}
      \hline\hline
      Parameter & Value\\\hline
      penalty & L1\\
      C & 1.668\\
      \hline\hline
    \end{tabular}
  \end{table}

  \begin{table}[htbp]
    \caption{\label{tab:hyperparameters-xgb-askap12} XGB hyperparameters for the `ASKAP' dataset.}
    \begin{tabular}{ll}
      \hline\hline
      Parameter & Value\\\hline
      colsample\textunderscore{}bytree & 0.865\\
      gamma & 0.256\\
      learning\textunderscore{}rate & 0.1\\
      max\textunderscore{}depth & 6\\
      min\textunderscore{}child\textunderscore{}weight & 1\\
      scale\textunderscore{}pos\textunderscore{}weight & 1\\
      subsample & 0.819\\
      n\textunderscore{}estimators & 108\\
      reg\textunderscore{}alpha & 0.049\\
      reg\textunderscore{}lambda & 0.454\\
      \hline\hline
    \end{tabular}
  \end{table}

  \begin{table}[htbp]
    \caption{\label{tab:hyperparameters-lr-askap12} LR hyperparameters for the `ASKAP' dataset.}
    \begin{tabular}{ll}
      \hline\hline
      Parameter & Value\\\hline
      penalty & L2\\
      C & 0.464\\
      \hline\hline
    \end{tabular}
  \end{table}

\section{Predictions on real data}
\label{sec:real-data-fig}

\begin{figure*}
    \centering
    \includegraphics[width=\linewidth]{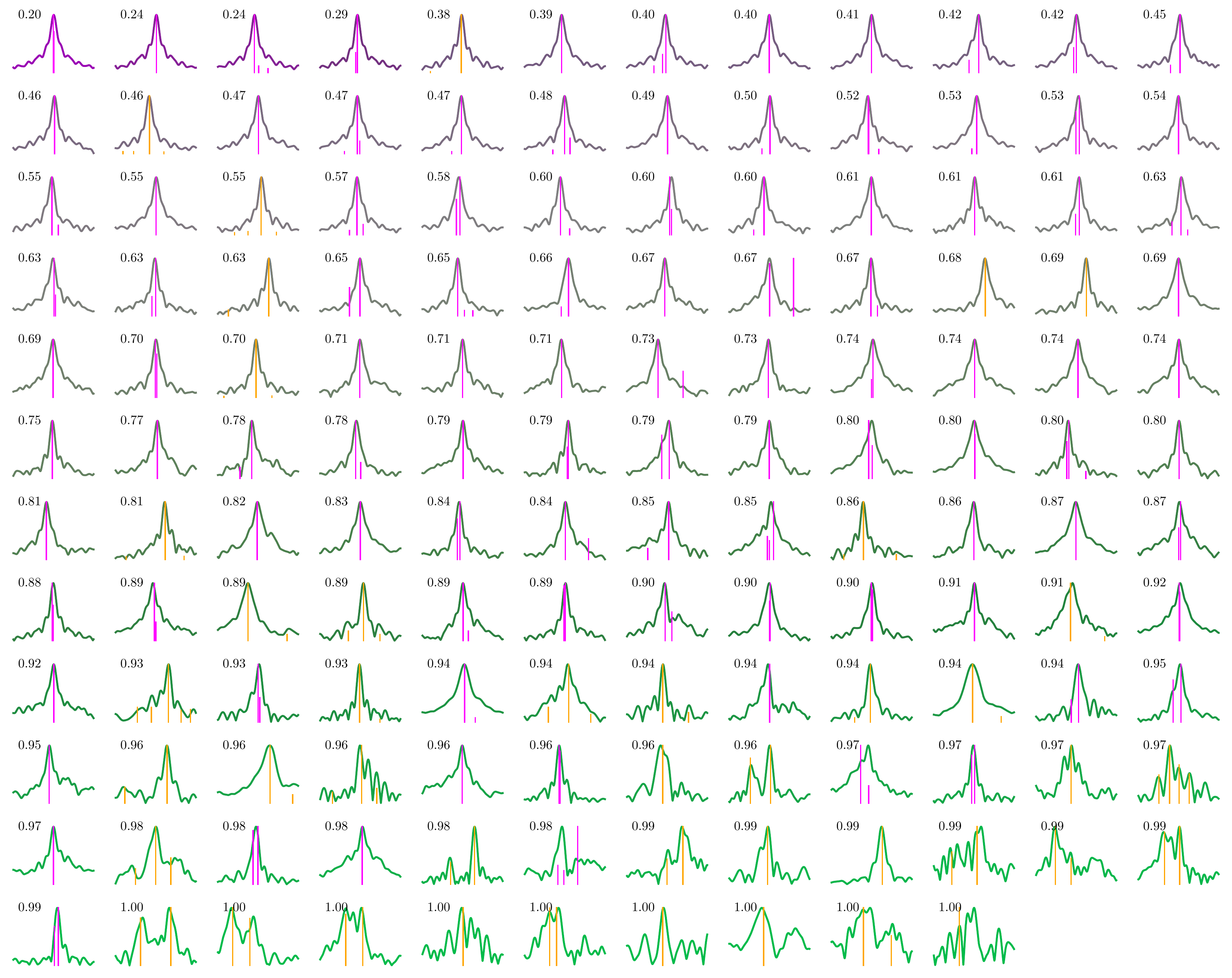}
    \caption{The 142 observed FDFs ordered by LR-estimated probability of being Faraday complex. Livingston-identified components are shown in orange while O'Sullivan-identified components are shown in magenta. Simpler FDFs (as deemed by the classifier) are shown in purple while more complex FDFs are shown in green, and the numbers overlaid indicate the LR estimate. A lower number indicates a lower probability that the corresponding source is complex, i.e. lower numbers correspond to simpler spectra.}
    \label{fig:all-observed-fdfs-lr}
  \end{figure*}

\begin{figure*}
    \centering
    \includegraphics[width=\linewidth]{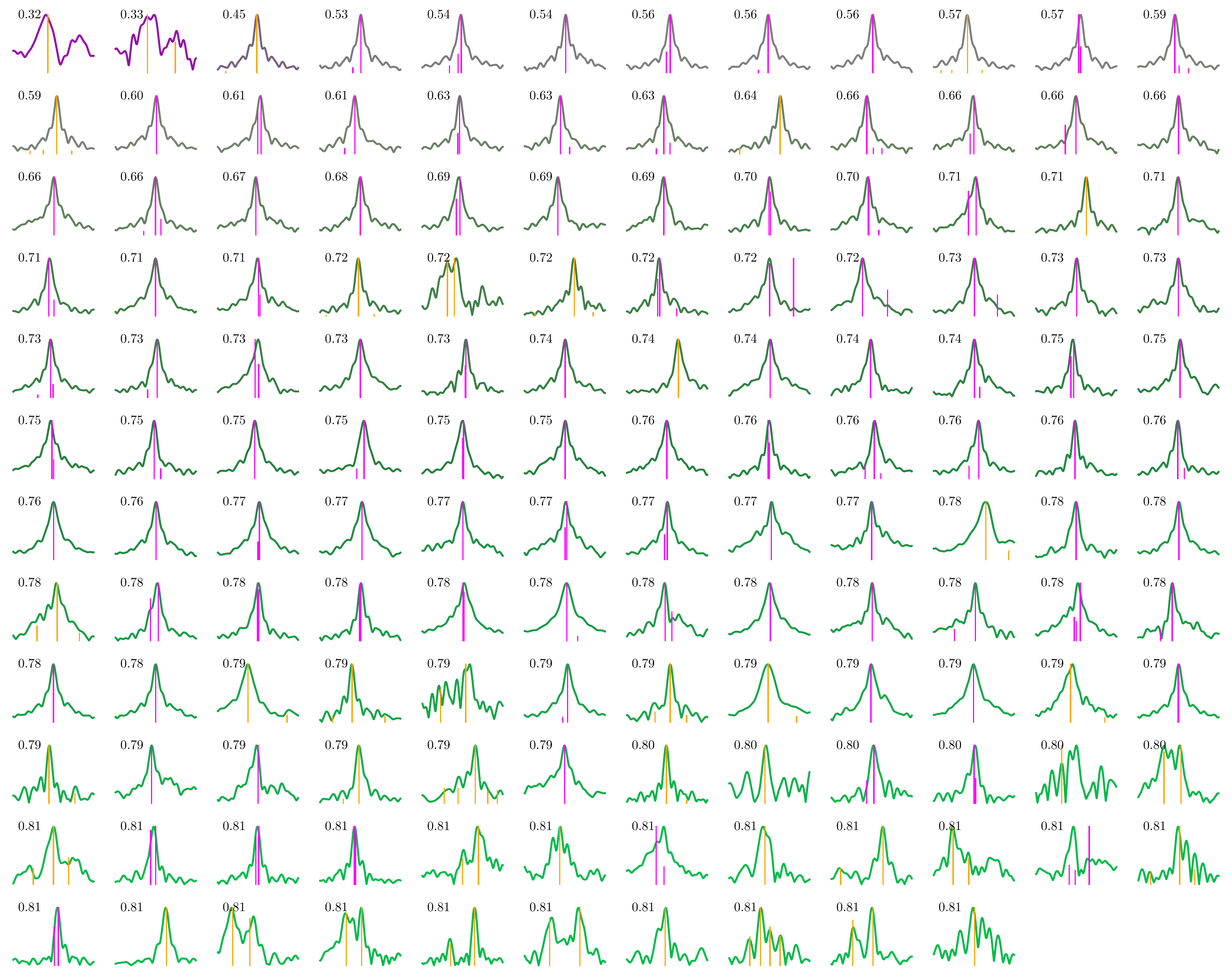}
    \caption{The 142 observed FDFs ordered by XGB-estimated probability of being Faraday complex. Livingston-identified components are shown in orange while O'Sullivan-identified components are shown in magenta. Simpler FDFs (as deemed by the classifier) are shown in purple while more complex FDFs are shown in green, and the numbers overlaid indicate the XGB estimate. A lower number indicates a lower probability that the corresponding source is complex, i.e. lower numbers correspond to simpler spectra.}
    \label{fig:all-observed-fdfs-xgb}
  \end{figure*}

  This section contains \autoref{fig:all-observed-fdfs-lr} and \autoref{fig:all-observed-fdfs-xgb}, which shows the predicted probability of being Faraday complex for all real data used in this paper, drawn from \citet{livingston21faraday} and \citet{osullivan_broad-band_2017}.

\section{Simulating observed FDFs}
\label{sec:simulating}

  We simulated FDFs by approximating them by arrays of complex numbers. An FDF $F$ is approximated on the domain $[-\phi_{\max}, \phi_{\max}]$ by a vector $\vec F \in \mathbb R^d$:
    \begin{equation}
      \label{eq:vec-f}
      \vec F_j = \sum_{k = 0}^1 A_k \delta(-\phi_{\max} + j \delta \phi - \phi_k)
    \end{equation}
    where $\delta\phi = (\phi_{\max} - \phi_{\min}) / d$ and $d$ is the number of Faraday depth samples in the FDF.
    $\vec F$ is sampled by uniformly sampling its parameters:
    \begin{align}
      \label{eq:model-distributions}
      \phi_k &\in [\phi_{\min}, \phi_{\min} + \delta\phi, \dots, \phi_{\max}]\\
      A_k &\sim \mathcal U(0, 1).
    \end{align}
    We then generate a vector polarisation spectrum $\vec P \in \mathbb R^m$ from $\vec F$ using a \autoref{eq:discrete-f-to-p}:
    \begin{equation}
      \label{eq:discrete-f-to-p}
      \vec P_\ell = \sum_{j = 0}^{j} F_j e^{2i(\phi_{\min} + j\delta_\phi)\lambda^2_\ell}\ \mathrm{d}\phi.
    \end{equation}
    $\lambda^2_\ell$ is the discretised value of $\lambda^2$ at the $\ell$th index of $\vec P$. This requires a set of $\lambda^2$ values, which depends on the dataset being simulated. These values can be treated as the channel wavelengths at which the polarisation spectrum was observed. We then add Gaussian noise with variance $\sigma^2$ to each element of $\vec P$ to obtain a discretised noisy observation $\hat{\vec{P}}$. Finally, we perform RM synthesis using the Canadian Initiative for Radio Astronomy Data Analysis \texttt{RM} package\footnote{\url{https://github.com/CIRADA-Tools/RM}}, which is a \texttt{Python} module that implements a discrete version of RM synthesis:
    \begin{equation}
      \label{eq:discrete-rm-synthesis}
      \hat{\vec{F}}_j = m^{-1} \sum_{\ell = 1}^m \vec{\hat P}_\ell e^{-2i(\phi_{\min} + j\delta_\phi)\lambda^2_\ell}.
    \end{equation}

\end{appendix}

\bibliographystyle{pasa-mnras}
\bibliography{sugar-glider}

\begin{thebibliography}{}
\makeatletter
\relax
\def\mn@urlcharsother{\let\do\@makeother \do\$\do\&\do\#\do\^\do\_\do\%\do\~}
\definecolor{darkblue}{rgb}{0,0,0.597656}
\def\mndoi{\begingroup\mn@urlcharsother \@ifnextchar [ {\mndoi@} {\mndoi@[]}}
\def\mndoi@[#1]#2{\def\@tempa{#1}\ifx\@tempa\@empty \href
  {http://dx.doi.org/#2} {\textcolor{darkblue}{doi:#2}}\else \href
  {http://dx.doi.org/#2} {\textcolor{darkblue}{#1}}\fi \endgroup}
\def\mn@eprint#1#2{\mn@eprint@#1:#2::\@nil}
\def\mn@eprint@arXiv#1{\href {http://arxiv.org/abs/#1} {{\tt arXiv:#1}}}
\def\mn@eprint@dblp#1{\href {http://dblp.uni-trier.de/rec/bibtex/#1.xml}
  {dblp:#1}}
\def\mn@eprint@#1:#2:#3:#4\@nil{\def\@tempa {#1}\def\@tempb {#2}\def\@tempc
  {#3}\ifx \@tempc \@empty \let \@tempc \@tempb \let \@tempb \@tempa \fi \ifx
  \@tempb \@empty \def\@tempb {arXiv}\fi \@ifundefined
  {mn@eprint@\@tempb}{\@tempb:\@tempc}{\expandafter \expandafter \csname
  mn@eprint@\@tempb\endcsname \expandafter{\@tempc}}}

\bibitem[\protect\citeauthoryear{Agarwal, Aggarwal, Burke-Spolaor, Lorimer  \&
  Garver-Daniels}{Agarwal et~al.}{2020}]{agarwal20fetch}
Agarwal D.,  Aggarwal K.,  Burke-Spolaor S.,  Lorimer D.~R.,   Garver-Daniels
  N.,  2020, \mndoi [\mnras] {10.1093/mnras/staa1856}

\bibitem[\protect\citeauthoryear{Anderson, Gaensler, Feain  \&
  Franzen}{Anderson et~al.}{2015}]{anderson_broadband_2015}
Anderson C.~S.,  Gaensler B.~M.,  Feain I.~J.,   Franzen T. M.~O.,  2015,
  \mndoi [\apj] {10.1088/0004-637X/815/1/49}, 815, 49

\bibitem[\protect\citeauthoryear{Brentjens \& de Bruyn}{Brentjens \&
  de~Bruyn}{2005}]{brentjens_faraday_2005}
Brentjens M.~A.,  de Bruyn A.~G.,  2005, \aap, 441, 1217

\bibitem[\protect\citeauthoryear{Brown}{Brown}{2011}]{Brown11report}
Brown S.,  2011, Assess the Complexity of an RM Synthesis Spectrum.
No.~9 in POSSUM REPORT

\bibitem[\protect\citeauthoryear{Brown et~al.,}{Brown
  et~al.}{2018}]{brown_classifying_2018}
Brown S.,  et~al., 2018, \mndoi [\mnras] {10.1093/mnras/sty2908}

\bibitem[\protect\citeauthoryear{{Farnes}, {Gaensler}  \& {Carretti}}{{Farnes}
  et~al.}{2014}]{farnes14broadband}
{Farnes} J.~S.,  {Gaensler} B.~M.,   {Carretti} E.,  2014, \mndoi [\apjs]
  {10.1088/0067-0049/212/1/15}, \href
  {https://ui.adsabs.harvard.edu/abs/2014ApJS..212...15F} {212, 15}

\bibitem[\protect\citeauthoryear{Flamary \& Courty}{Flamary \&
  Courty}{2017}]{flamary17pot}
Flamary R.,  Courty N.,  2017, POT Python Optimal Transport library, \url
  {https://github.com/rflamary/POT}

\bibitem[\protect\citeauthoryear{Gebhard, Kilbertus, Harry  \&
  Sch\"olkopf}{Gebhard et~al.}{2019}]{gebhard19convolutional}
Gebhard T.~D.,  Kilbertus N.,  Harry I.,   Sch\"olkopf B.,  2019, \mndoi [\prd]
  {10.1103/PhysRevD.100.063015}, 100, 063015

\bibitem[\protect\citeauthoryear{{Goldstein} \& {Reed}}{{Goldstein} \&
  {Reed}}{1984}]{goldstein84faraday}
{Goldstein} S.~J. J.,  {Reed} J.~A.,  1984, \mndoi [\apj] {10.1086/162337},
  \href {https://ui.adsabs.harvard.edu/abs/1984ApJ...283..540G} {283, 540}

\bibitem[\protect\citeauthoryear{Heald}{Heald}{2008}]{heald09faraday}
Heald G.,  2008, \mndoi [Proceedings of the International Astronomical Union]
  {10.1017/S1743921309031421}, 4, 591

\bibitem[\protect\citeauthoryear{Hlo\v{z}ek et~al.,}{Hlo\v{z}ek
  et~al.}{2020}]{hlozek20lsst}
Hlo\v{z}ek R.,  et~al., 2020, arXiv e-prints, 2012, arXiv:2012.12392

\bibitem[\protect\citeauthoryear{{Law} et~al.,}{{Law}
  et~al.}{2011}]{law11faraday}
{Law} C.~J.,  et~al., 2011, \mndoi [\apj] {10.1088/0004-637X/728/1/57}, \href
  {https://ui.adsabs.harvard.edu/abs/2011ApJ...728...57L} {728, 57}

\bibitem[\protect\citeauthoryear{Livingston, McClure-Griffiths, Gaensler, Seta
  \& Alger}{Livingston et~al.}{2021}]{livingston21faraday}
Livingston J.~D.,  McClure-Griffiths N.~M.,  Gaensler B.~M.,  Seta A.,   Alger
  M.~J.,  2021, \mndoi [\mnras] {10.1093/mnras/stab253}

\bibitem[\protect\citeauthoryear{Ma, Mao, Stil, Basu, West, Heiles, Hill  \&
  Betti}{Ma et~al.}{2019}]{ma_broad-band_2019}
Ma Y.~K.,  Mao S.~A.,  Stil J.,  Basu A.,  West J.,  Heiles C.,  Hill A.~S.,
  Betti S.~K.,  2019, \mndoi [\mnras] {10.1093/mnras/stz1325}, 487, 3432

\bibitem[\protect\citeauthoryear{Machado Poletti~Valle, Avestruz, Barnes,
  Farahi, Lau  \& Nagai}{Machado Poletti~Valle et~al.}{2020}]{valle20shap}
Machado Poletti~Valle L.~F.,  Avestruz C.,  Barnes D.~J.,  Farahi A.,  Lau
  E.~T.,   Nagai D.,  2020, arXiv e-prints, 2011, arXiv:2011.12987

\bibitem[\protect\citeauthoryear{Miyashita, Ideguchi, Nakagawa, Akahori  \&
  Takahashi}{Miyashita et~al.}{2019}]{miyashita19qu}
Miyashita Y.,  Ideguchi S.,  Nakagawa S.,  Akahori T.,   Takahashi K.,  2019,
  \mndoi [\mnras] {10.1093/mnras/sty2862}, 482, 2739

\bibitem[\protect\citeauthoryear{{O'Sullivan} et~al.,}{{O'Sullivan}
  et~al.}{2012}]{osullivan12agn}
{O'Sullivan} S.~P.,  et~al., 2012, \mndoi [\mnras]
  {10.1111/j.1365-2966.2012.20554.x}, \href
  {https://ui.adsabs.harvard.edu/abs/2012MNRAS.421.3300O} {421, 3300}

\bibitem[\protect\citeauthoryear{O'Sullivan, Purcell, Anderson, Farnes, Sun  \&
  Gaensler}{O'Sullivan et~al.}{2017}]{osullivan_broad-band_2017}
O'Sullivan S.~P.,  Purcell C.~R.,  Anderson C.~S.,  Farnes J.~S.,  Sun X.~H.,
  Gaensler B.~M.,  2017, \mndoi [\mnras] {10.1093/mnras/stx1133}, 469, 4034

\bibitem[\protect\citeauthoryear{{Pan} \& {Yang}}{{Pan} \&
  {Yang}}{2010}]{pan10transfer}
{Pan} S.~J.,  {Yang} Q.,  2010, \mndoi [IEEE Transactions on Knowledge and Data
  Engineering] {10.1109/TKDE.2009.191}, 22, 1345

\bibitem[\protect\citeauthoryear{Sun et~al.,}{Sun
  et~al.}{2015}]{sun15comparison}
Sun X.~H.,  et~al., 2015, \mndoi [The Astronomical Journal]
  {10.1088/0004-6256/149/2/60}, 149, 60

\bibitem[\protect\citeauthoryear{Van~Eck et~al.,}{Van~Eck
  et~al.}{2017}]{vaneck17faraday}
Van~Eck C.~L.,  et~al., 2017, \mndoi [\aap] {10.1051/0004-6361/201629707}, 597,
  A98

\bibitem[\protect\citeauthoryear{{Virtanen} et~al.,}{{Virtanen}
  et~al.}{2020}]{scipy2020}
{Virtanen} P.,  et~al., 2020, \mndoi [Nature Methods]
  {https://doi.org/10.1038/s41592-019-0686-2}, \href {https://rdcu.be/b08Wh}
  {17, 261}

\bibitem[\protect\citeauthoryear{Zevin et~al.,}{Zevin
  et~al.}{2017}]{zevin17gravityspy}
Zevin M.,  et~al., 2017, \mndoi [Classical and Quantum Gravity]
  {10.1088/1361-6382/aa5cea}, 34, 064003

\bibitem[\protect\citeauthoryear{Zhang}{Zhang}{2019}]{zhang2019transfer}
Zhang L.,  2019, arXiv preprint arXiv:1903.04687

\bibitem[\protect\citeauthoryear{Zhu, Ong  \& Huttley}{Zhu
  et~al.}{2020}]{zhu20mutagenic}
Zhu Y.,  Ong C.~S.,   Huttley G.~A.,  2020, \mndoi [Genetics]
  {10.1534/genetics.120.303093}, 215, 25

\makeatother
\end{thebibliography}

\end{document}